\documentclass[times,twocolumn]{aastex6}
\usepackage[fleqn]{amsmath}
\usepackage{natbib}
\usepackage{enumitem}
\usepackage{booktabs}
\usepackage{graphicx}
\usepackage{float}
\usepackage[caption=false]{subfig}

\begin{document}

\title{Recalibration of the $M_{BH}-\sigma_{\star}$ Relation for AGN}

\author{Merida Batiste\altaffilmark{1}, Misty C. Bentz\altaffilmark{1}, Sandra I. Raimundo\altaffilmark{2}, Marianne Vestergaard\altaffilmark{2,3}, Christopher A. Onken\altaffilmark{4}}
\begin{center}
\altaffiltext{1}{Department of Physics \& Astronomy, Georgia State University, 25 Park Place, Atlanta, GA 30303, USA; batiste@astro.gsu.edu}
\altaffiltext{2}{Dark Cosmology Centre, Niels Bohr Institute, University of Copenhagen, Juliane Maries Vej 30, DK-2100 Copenhagen {{\O}}, Denmark}
\altaffiltext{3}{Department of Astronomy and Steward Observatory, University of Arizona, 933 N. Cherry Avenue, Tucson, AZ 85721, USA}
\altaffiltext{4}{Research School of Astronomy \& Astrophysics, The Australian National University, Canberra, ACT 2611, Australia}

\end{center}

\begin{abstract}
We present a re-calibration of the $M_{BH}-\sigma_{\star}$ relation, based on a sample of 16 reverberation-mapped galaxies with newly determined bulge stellar velocity dispersions ($\sigma_{\star}$) from integral-field  spectroscopy (IFS), and a sample of 32 quiescent galaxies with publicly available IFS. For both samples, $\sigma_{\star}$ is determined via two different methods that are popular in the literature, and we provide fits for each sample based on both sets of $\sigma_{\star}$. We find the fit to the AGN sample is shallower than the fit to the quiescent galaxy sample, and that the slopes for each sample are in agreement with previous investigations. However, the intercepts to the quiescent galaxy relations are notably higher than those found in previous studies, due to the systematically lower $\sigma_{\star}$ measurements that we obtain from IFS. We find that this may be driven, in part, by poorly constrained measurements of bulge effective radius ($r_{e}$) for the quiescent galaxy sample, which may bias the $\sigma_{\star}$ measurements low. We use these quiescent galaxy parameterizations, as well as one from the literature, to recalculate the virial scaling factor $f$. We assess the potential biases in each measurement, and suggest $f=4.82\pm1.67$ as the best currently available estimate. However, we caution that the details of how $\sigma_{\star}$ is measured can significantly affect $f$, and there is still much room for improvement.

\end{abstract}

\section{Introduction} \label{sec:intro}
A wealth of evidence demonstrates that the formation and evolution of galaxies and their supermassive black holes (BHs) are fundamentally connected. This connection is exemplified by empirically determined scaling relations between the mass of a central BH, $M_{BH}$, and host galaxy properties, including bulge stellar velocity dispersion, $\sigma_{\star}$ \citep{Ferrarese00,Gebhardt00}. These scaling relations provide insight into the mechanisms governing the formation and evolution of structure, and may be used to estimate $M_{BH}$ for large samples of galaxies at cosmological distances. %Thus they are essential tools in probing the evolution of structure across cosmic time. 

Accurate calibration of scaling relations requires a sample of galaxies with secure $M_{BH}$ determinations. In quiescent galaxies this is usually done by modeling the spatially resolved gas or stellar kinematics within the gravitational sphere of influence of the BH, and is thus limited to the local universe. To probe $M_{BH}$ over cosmological distances requires active galactic nuclei (AGNs), for which $M_{BH}$ can be determined via reverberation-mapping (RM) \citep{Blandford82}. RM exploits the variability of the AGN to probe the gas in the broad-line region (BLR) around the BH. A dimensionless scale factor $f$ is necessary for this method, to account for the unknown geometry and kinematics of the BLR. Since direct determination of $f$ is rarely feasible, it is assumed that AGN and quiescent galaxies follow the same $M_{BH}-\sigma_{\star}$ relation. The value of $f$ is then estimated as the average multiplicative offset required to bring the relations for AGN and quiescent galaxies into agreement \citep{Onken04}. Accurate calibration of the $M_{BH}-\sigma_{\star}$ relation is, therefore, essential for RM $M_{BH}$ determinations.
 
The $M_{BH}-\sigma_{\star}$ relation appears to be the tightest and most fundamental of the observed scaling relations (e.g. \citealt{Beifiori12,Shankar16}), and has consequently been the subject of extensive investigation (see reviews by \citealt{Kormendy13}, hereafter KH13, and \citealt{Graham16}). However it remains unclear what the actual best-fitting relation is, or indeed whether a single relation holds for both active and quiescent galaxies. Studies suggest a significant difference between the slopes of the relation for quiescent galaxies \citep{Mcconnell13} and AGN \citep{Woo10}, %which may critically impact the scaling of all RM $M_{BH}$, 
however simulations indicate that this may simply be an artifact of sample selection bias (\citealt{Shankar16,Woo13}, hereafter W13). 

Studies further indicate a possible morphological dependence of the $M_{BH}-\sigma_{\star}$ relation. In particular, that galaxies with substructure such as bars and pseudo-bulges are offset from the elliptical-only relation (e.g. \citealt{Graham08,Hu08,Gultekin09}). This is particularly relevant when measuring $\sigma_{\star}$ for AGN, which is often done via single-aperture and long-slit spectroscopy. Contamination by dynamically distinct substructure is usually unavoidable, and rotational broadening due to disk contamination can strongly affect $\sigma_{\star}$ measurements from single-aperture spectra (e.g. \citealt{Graham11,Hartmann14,Bellovary14}, W13). In addition, \cite{Batiste16} (hereafter B17) showed that slit orientation relative to substructure, such as bars, can strongly affect the measured $\sigma_{\star}$. These issues preferentially impact the spiral dominated local RM AGN sample, thereby inhibiting investigation of possible differences between quiescent and active galaxies. 

This problem is mitigated by spatially resolved kinematics from integral-field spectroscopy (IFS), which allows for significant improvement in $\sigma_{\star}$ determinations% for both the RM AGN sample and the quiescent galaxy sample
. B17 provide IFS-based $\sigma_{\star}$ estimates for ten RM AGN, and IFS is available in the literature for a further six. IFS is also available for 32 quiescent galaxies with dynamical $M_{BH}$ measurements. 
In this letter we use these samples to re-calibrate the $M_{BH}-\sigma_{\star}$ relation for quiescent galaxies and AGN, and provide a new estimate of the scale factor $f$ for use with RM $M_{BH}$ determination. Throughout this work we adopt a $\Lambda$CDM cosmology with $\Omega_{m}=0.3$, $\Omega_{\Lambda}=0.7$, and $H_{0}=70\:\mathrm{km\:s^{-1}\:Mpc^{-1}}$.

\section{The AGN Sample} \label{sec:AGN}
We include in our analysis all 16 RM AGN host galaxies which have so far been observed with IFS. While this is a small subset of the full RM AGN sample, the rest of which are targets of an ongoing observational campaign, it does provide a representative overview of the full sample for the $\sigma_{\star}$ and $M_{BH}$ ranges probed.
\subsection{Virial Products from Reverberation-Mapping}\label{subsec:VPs}
RM allows accurate determination of the virial product (VP), given by $\mathrm{VP}=\:V^{2}R_{BLR}/G$, where G is the gravitational constant, $V$ is measured from the width of a broad emission line, and $R_{BLR}$ is the size of the BLR. VPs are drawn from the AGN Black Hole Mass Database\footnote{http://www.astro.gsu.edu/AGNmass/} \citep{Bentz15} and from \cite{Bentz16a}, and are listed in column (7) of Table \ref{tab:data}. In all cases, VP is determined from the $\mathrm{H} \beta$ line. Individual references are available from the database.

\subsection{Bulge Stellar Velocity Dispersions}\label{subsec:kin_map}

Spatially resolved stellar kinematics are available for NGC 5273, from the ATLAS$^{3D}$ survey of early-type galaxies in the northern hemisphere\footnote{http://www-astro.physics.ox.ac.uk/atlas3d/} \citep{Cappellari11,Cappellari13}, and for MGC-06-30-15, from the work of \cite{Raimundo13}. 
Following the method of B17, $\sigma_{\star}$ is determined for these galaxies by taking an error-weighted average of the values for each spaxel within a circular aperture defined by the effective radius, $r_{e}$.

Accurate measurements of $r_{e}$ are available from the works of \cite{Bentz14} for NGC 5273, and \cite{Bentz16a} for MGC-06-30-15. They are determined from detailed surface brightness decompositions of Hubble Space Telescope (HST) images, using GALFIT \citep{Peng02,Peng10}. The AGN is isolated from the galaxy surface brightness features, and substructure such as bars and disks are accounted for.

\begin{figure}
%\begin{minipage}{\textwidth}
\begin{center}
\includegraphics[scale=1.05]{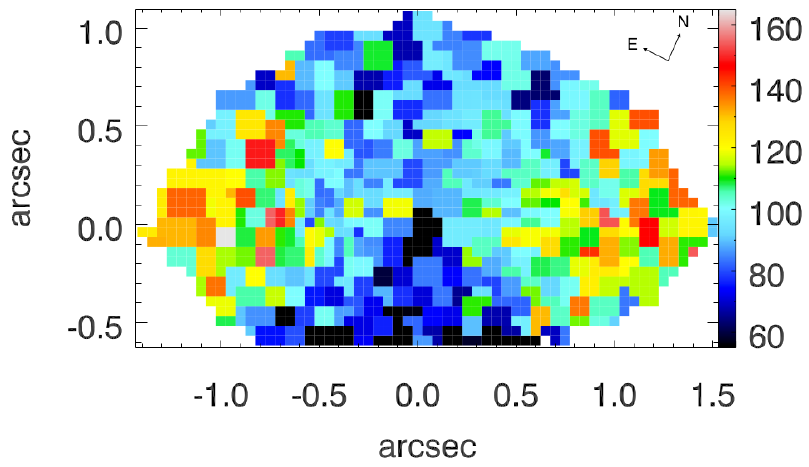}
\end{center}
\caption{Map of $\sigma_{\star}$ for MCG-06-30-15, based on data from \cite{Raimundo13}.}
%\end{minipage}
\label{fig:MCG63015}
\end{figure}

For MCG-06-30-15, we follow \cite{Raimundo13} and exclude the central $0\farcs1$ from our calculation of $\sigma_{\star}$, as the noise associated with the AGN continuum precludes secure measurement of the stellar kinematics. Furthermore, the kinematic map in Figure \ref{fig:MCG63015} shows that a small region within $r_{e}$ ($1\farcs01$) was cut off ($y\leq-0\farcs6$, the bottom of the map), due to an illumination artifact in the SINFONI data (see \citealt{Raimundo13} for details).
We assume that the kinematics within $r_{e}$ are well represented by the region that remains, and determine $\sigma_{\star}$ without applying any correction. 

Two estimates of uncertainty are provided for each $\sigma_{\star}$ determination. The statistical uncertainty, based on the measurement error, is shown in column (4) of Table \ref{tab:data} along with $\sigma_{\star}$. Column (6) shows the standard deviation among the set of $\sigma_{\star}$ values that have been averaged to determine the overall $\sigma_{\star}$. The standard deviation provides a measure of the spatial variation in the kinematics within $r_{e}$, and may be more physically meaningful as an estimate of uncertainty.

Finally, $\sigma_{\star}$ determinations from IFS are available for 4 high-luminosity quasar hosts from \cite{Grier13}. This study employs a different definition of $\sigma_{\star}$, which includes a contribution from the rotational velocity. Rather than averaging the kinematics within a chosen aperture, the spectra within that aperture are instead co-added, and $\sigma_{\star}$ is measured from the resulting rotationally broadened spectrum. While this method differs from that of B17, it has been favored in some recent studies, including those using IFS (e.g. \citealt{Gultekin09,Cappellari13,Vandenbosch16}, KH13). For elliptical or near face-on disk galaxies the difference between the methods should be minimal, since rotation along the line-of-sight will not dominate the stellar kinematics (W13). Based on the GALFIT decompositions of \cite{Bentz09}, two of these quasar hosts are elliptical (i.e. fitted with only a bulge component), and two are low-inclination disk galaxies. Consequently we do not expect any bias to arise from including these measurements.

We can directly test this expectation because our sample contains galaxies qualitatively similar to the quasar hosts, including the low-inclination disk galaxies NGC 5273 and NGC 6814. We measure $\sigma_{\star}$ for the whole sample via the method of \cite{Grier13} (column (5) of Table \ref{tab:data}). The greatest variation between $\sigma_{\star}$ estimates from the two methods occurs for inclined spiral galaxies with significant substructure (e.g. NGC 4593) or evidence of ongoing or recent interactions (e.g. NGC 3227, MCG-06-30-15, and NGC 5548, \citealt{Bentz16}). As expected, $\sigma_{\star}$ varies minimally between the two methods for NGC 5273 and NGC 6814, suggesting that including the quasars is unlikely to introduce significant bias.

Measurements of $\sigma_{\star}$ for the full sample are given in Table \ref{tab:data}.

\begin{table*}
\begin{minipage}{\textwidth}
\begin{center}
\caption{AGN sample}
\label{tab:data}
\begin{tabular}{lccccccccc}
\hline
\hline
Object & $r_{e}$ & Ref.  &$\sigma_{\star}$	&$\sigma_{\star_{int}}$&std deviation & Ref. & VP & Morphological Type\\
  & ($\arcsec$) & 	& $(\mathrm{km\:s^{-1}}) $ & $\mathrm{(km\:s^{-1})}$& $\mathrm{(km\:s^{-1})}$ & & $(10^{7}\:M_{\odot})$ & \\
  (1) & (2) & (3) & (4) & (5) & (6) & (7) & (8) & (9)\\
\hline				  
Mrk 79 & 2.0 & 1 &$120\pm9	$&$125\pm15	$& 21&6 &$0.951^{+0.267}_{-0.256}$& barred late\\	  
NGC 3227	& 2.7 & 5 &$114\pm3	$&$136\pm6	$& 13	&6 &$0.139^{+0.029}_{-0.032}$&barred late\\	  
NGC 3516	& 2.1 & 5 &$139\pm4	$&$143\pm4	$& 12	&6 &$0.577^{+0.051}_{-0.076}$&barred late\\	  
NGC 4051	& 1.0 & 5 &$74 \pm2	$&$69 \pm4	$& 4 	&6 &$0.031^{+0.010}_{-0.009}$&barred late\\	  
NGC 4151	& 2.1 & 5 &$105\pm5	$&$110\pm8	$& 15	&6 &$0.923^{+0.163}_{-0.115}$&barred late\\  
NGC 4253	& 1.4 & 2 &$84 \pm4	$&$85 \pm9	$& 9 	&6 &$0.032^{+0.028}_{-0.028}$&barred late\\  
NGC 4593	& 11.5 & 5 &$113\pm3	$&$144\pm5	$& 14	&6 &$0.177^{+0.038}_{-0.038}$&barred late\\
NGC 5273& 6.8 & 3 &$62 \pm3	$&$69 \pm5	$&  9 	&8 &$0.103^{+0.057}_{-0.076}$&early\\  	
Mrk 279& 1.6 & 1 &$153\pm7	$&$156\pm17	$& 26	&6 &$0.657^{+0.177}_{-0.177}$&early\\  
PG 1411+442& 3.1 & 1 & --&$208\pm30$& -- & 7 &$6.263^{+3.344}_{-3.376}$&early\\  
NGC 5548& 11.2 & 5 &$131\pm3	$&$162\pm12	$& 34	&6 &$1.212^{+0.052}_{-0.050}$&late\\  
PG 1617+175& 1.7 & 1 & -- &$201\pm37$& -- & 7 &$9.620^{+4.272}_{-4.790}$&early\\  
NGC 6814& 1.7 & 2 &$71 \pm3	$&$69 \pm3	$& 5 	&6 &$0.336^{+0.063}_{-0.064}$&barred late\\  
Mrk 509& 2.8 & 1 & -- &$183\pm12$ & -- & 7 &$2.529^{+0.223}_{-0.204}$&early\\  
PG 2130+099& 0.32 & 1 & -- &$165\pm19$& -- &7 &$0.630^{+0.086}_{-0.086}$&late\\   
MCG-06-30-15& 1.01 & 4 &$95\pm5$&$91\pm5$& 22&8 &$0.037^{+0.010 }_{-0.009 }$&early\\
\hline
\end{tabular}
\end{center}
\textbf{References.} (1) \citealt{Bentz09}; (2) \citealt{Bentz13}; (3) \citealt{Bentz14}; (4) \citealt{Bentz16a}; (5) \citealt{Bentz16}; (6) \citealt{Batiste16}; (7) \citealt{Grier13}; (8) this work\\
\textbf{Notes.} Column 1: galaxy name, Column 2: $r_{e}$, Column 3: reference for $r_{e}$, Column 4: $\sigma_{\star}$ within $r_{e}$ with associated $1\sigma$ uncertainty, Column 5: $\sigma_{\star}$ measured from a single spectrum integrated within a circular aperture of radius $r_{e}$, with associated $1\sigma$ uncertainty, Column 6: standard deviation for the set of $\sigma_{\star}$ values averaged to determine the value in column 4. No value is included for galaxies where an integrated spectrum was used, Column 7: reference for $\sigma_{\star}$, Column 8: VP, Column 9: morphological type, based on surface brightness decompositions of \cite{Bentz09,Bentz13,Bentz16}. 
%\vspace{2ex} 
\end{minipage}
\end{table*}

\section{The Quiescent Galaxy Sample}\label{sec:quiescent_sample}

\cite{Cappellari13} provide stellar kinematics for 32 quiescent galaxies from the compilation of KH13 (listed in Table \ref{tab:quiescent_data}), with $\sigma_{\star}$ determined via the same method as \cite{Grier13}. For comparison, we calculate $\sigma_{\star}$ for each galaxy following the method of B17, using $r_{e}$ measurements from \cite{Cappellari11}. The sample contains elliptical and disk galaxies, so these methods give quite different results in some cases. On average the measurements of \cite{Cappellari13} are larger, by $\sim13\:\mathrm{km\:s^{-1}}$. 

It is essential to note here that while the stellar kinematics are high quality, the measurements of $r_{e}$ are less reliable (see discussion by \citealt{Cappellari13}). Measurements come from seeing-limited ground-based images, rather than HST images, as are used for the AGN sample. Moreover, $r_{e}$ was defined as the radius containing half the observed light for the \textit{whole galaxy} \citep{Cappellari13}, rather than for the bulge. For the S0 galaxy NGC 5273, we find that the value quoted by \cite{Cappellari11} ($r_{e}=37\farcs15$) is more than $5\times$ larger than that determined from the bulge-disk decompositions of \cite{Bentz14}.

\cite{Falcon17} have shown that, for early-type galaxies (including lenticulars and Sa galaxies), $\sigma_{\star}$ generally decreases with radius. Consequently, if we assume that $r_{e}$ is typically over-estimated by $\sim5\times$ for the quiescent sample, then the corresponding $\sigma_{\star}$ measurements are likely biased low. However, for inclined disk galaxies the inclusion of disk rotation at large radii may bias the estimates high (e.g. \citealt{Bellovary14}). The quiescent sample contains a range of galaxy morphologies and disk inclinations, so there are multiple reasons to be cautious with the adopted $r_{e}$ values and the quoted $\sigma_{\star}$ measurements.

Since similar measurements of $r_{e}$ have been used in previous studies (e.g. \citealt{Gultekin09}, KH13), it is safe to assume that all studies of the $M_{BH}-\sigma_{\star}$ relation are affected by this issue to some extent, and substantially improved measurements of $r_{e}$ will be critical to all future efforts to properly calibrate the $M_{BH}-\sigma_{\star}$ relation.

\section{The $M_{BH}-\sigma_{\star}$ Relation}\label{sec:relation}
The $M_{BH}-\sigma_{\star}$ relation is parameterized as:
\begin{equation}\label{eq:Msigma}
\mathrm{log\:\left(\dfrac{M_{BH}}{M_{\odot}}\right)} = \alpha\: +\: \beta\:\mathrm{log}\left(\dfrac{\sigma_{\star}}{200\:\mathrm{km\:s^{-1}}}\right)
\end{equation}
We fit a standard forward regression using the LINMIX\_ERR routine of \cite{Kelly07}, which employs a fully Bayesian approach. We also tested the popular MPFITEXY routine of \cite{Williams10}, since \cite{Park12} showed that both are similarly robust and unbiased, and found that the results are consistent with those determined by LINMIX\_ERR. 

\subsection{The Quiescent Galaxy Sample}\label{subsec:quiescent_msig}
Using $\sigma_{\star}$ from \cite{Cappellari13} we find a best fitting relation for the quiescent sample of:
\begin{equation}\label{eq:Msigma_quiescent}
\mathrm{log\:\left(\dfrac{M_{BH}}{M_{\odot}}\right)} = (8.55\pm0.09)\: +\: (5.32\pm0.63)\:\mathrm{log}\left(\dfrac{\sigma_{\star}}{200\:\mathrm{km\:s^{-1}}}\right)
\end{equation}
This agrees remarkably well with the parameterization of W13, who find a slope of $5.31\pm0.33$, and is consistent with that of \cite{Grier13} ($5.04\pm0.19$), and that of \cite{Savorgnan15} ($6.34\pm0.8$).

We find a slightly shallower slope when we use our own determinations of $\sigma_{\star}$, more consistent with that of KH13 ($4.38\pm0.29$):
\begin{equation}\label{eq:Msigma_quiescent2}
\mathrm{log\:\left(\dfrac{M_{BH}}{M_{\odot}}\right)} = (8.66\pm0.09)\: +\: (4.76\pm0.60)\:\mathrm{log}\left(\dfrac{\sigma_{\star}}{200\:\mathrm{km\:s^{-1}}}\right)
\end{equation}
%though the slopes from Equations \ref{eq:Msigma_quiescent} and \ref{eq:Msigma_quiescent2} agree within uncertainties. 

While the slopes are consistent with the literature, the intercepts are higher. Comparing Equation \ref{eq:Msigma_quiescent} with the parameterization by W13 (who find $\alpha=8.37\pm0.05$) is particularly instructive, since the slopes are almost identical. Differences between the intercepts arise from systematic differences between the sample of $\sigma_{\star}$ measurements. On average, the $\sigma_{\star}$ measurements from Atlas$^{3D}$ data are lower than the literature values, causing the relation to be shifted left, thus increasing the intercept. While lower $\sigma_{\star}$ values are expected from IFS (see e.g. KH13, B17), these measurements may be biased low if $r_{e}$ are over-estimated (Section \ref{sec:quiescent_sample}). Thus the intercepts that we measure are likely too high, while the intercepts quoted in the literature are probably too low.

\begin{table*}
\begin{minipage}{\textwidth}
\begin{center}
\caption{Quiescent Galaxies}
\label{tab:quiescent_data}
\begin{tabular}{lcccc}
\hline
\hline
Galaxy & $\sigma_{\star_{ATLAS}}$ &	$\sigma_{\star_{calc}}$& standard deviation  & $\mathrm{M_{BH}}$\\
       	 &  $(\mathrm{km\:s^{-1}})$ &	$(\mathrm{km\:s^{-1}})$ &$(\mathrm{km\:s^{-1}})$ & $(10^{7}\:M_{\odot})$\\
 (1) 	&	(2)	&	(3)	& (4)	&	(5)	\\
\hline
NGC 0524   &$220\pm	11$&  $206\pm	12$	&  $22$	&  $86.7^{+9.4}_{-4.6}$\\
NGC 0821   &$179\pm	9 $&  $173\pm	10$	&  $17$	&  $16.5^{+7.4}_{-7.3}$\\
NGC 1023   &$167\pm	8 $&  $145\pm	7 $	&  $34$	&  $4.1^{+0.4}_{-0.4 }$\\
NGC 2549   &$142\pm	7 $&  $109\pm	10$	&  $19$	&  $1.5^{+0.2}_{-1.1 }$\\
NGC 2778   &$132\pm	7 $&  $112\pm	11$	&  $28$	&  $1.5^{+1.5}_{-1.5 }$\\
NGC 3245   &$177\pm	9 $&  $126\pm	9 $	&  $30$	&  $23.9^{+2.7}_{-7.6}$\\
NGC 3377   &$128\pm	6 $&  $105\pm	10$	&  $19$	&  $17.8^{+9.4}_{-9.3}$\\
NGC 3379   &$186\pm	9 $&  $185\pm	9 $	&  $16$	&  $41.6^{+10.4}_{-10.4}$\\
NGC 3384   &$138\pm	7 $&  $118\pm	7 $	&  $19$	&  $1.1^{+0.5}_{-0.5   }$\\
NGC 3489   &$101\pm	5 $&  $74\pm	  	8 $	&  $14$	&  $0.6^{+0.1}_{-0.1}$\\
NGC 3607   &$207\pm	10$&  $206\pm	11$	&  $17$	&  $13.7^{+4.5}_{-4.7  }$\\
NGC 3608   &$169\pm	8 $&  $166\pm	10$	&  $17$	&  $46.5^{+9.9}_{-9.9  }$\\
NGC 3945   &$177\pm	9 $&  $141\pm	10$	&  $25$	&  $0.9^{+1.7}_{-0.9   }$\\
NGC 3998   &$224\pm	11$&  $182\pm	9 $	&  $34$	&  $84.5^{+7.0}_{-6.6  }$\\
NGC 4026   &$157\pm	8 $&  $123\pm	9 $	&  $23$	&  $18.0^{+6.0}_{-3.5  }$\\
NGC 4261   &$265\pm	13$&  $285\pm	11$	&  $22$	&  $52.9^{+10.7}_{-10.8}$\\
NGC 4342   &$242\pm	12$&  $192\pm	7 $	&  $29$	&  $45.3^{+26.5}_{-14.8}$\\
NGC 4374   &$258\pm	13$&  $271\pm	9 $	&  $20$	&  $92.5^{+9.8}_{-8.7}$\\
NGC 4382   &$179\pm	9 $&  $190\pm	7 $	&  $11$	&  $1.3^{+21.1}_{-1.3}$\\
NGC 4459   &$158\pm	8 $&  $135\pm	10$	&  $20$	&  $7.0^{+1.3}_{-1.3}$\\
NGC 4472   &$250\pm	13$&  $266\pm	7 $	&  $18$	&  $254.0^{+58.0}_{-10.0}$\\
NGC 4473   &$187\pm	9 $&  $176\pm	8 $	&  $22$	&  $9.0^{+4.5}_{-4.5}$\\
NGC 4486 (M87)   &$264\pm	13$&  $295\pm	4 $	&  $19$	&  $615.0^{+38.0}_{-37.0}$\\
NGC 4486A  &$123\pm	6 $&  $115\pm	16$	&  $71$	&  $1.4^{+0.5}_{-0.5}$\\
NGC 4526   &$209\pm	10$&  $175\pm	7 $	&  $30$	&  $45.1^{+14.0}_{-10.3}$\\
NGC 4564   &$155\pm	8 $&  $134\pm	9 $	&  $24$	&  $8.8^{+2.5}_{-2.4}$\\
NGC 4596   &$126\pm	6 $&  $127\pm	10$	&  $31$	&  $7.7^{+3.7}_{-3.2}$\\
NGC 4649   &$268\pm	13$&  $283\pm	7 $	&  $22$	&  $472.0^{+104.0}_{-105.0}$\\
NGC 4697   &$169\pm	8 $&  $166\pm	9 $	&  $13$	&  $20.2^{+5.1}_{-5.0}$\\
NGC 5576   &$155\pm	8 $&  $155\pm	10$	&  $22$	&  $27.3^{+6.8}_{-7.9}$\\
NGC 5845   &$228\pm	11$&  $178\pm	6 $	&  $35$	&  $48.7^{15.3}_{-15.3}$\\
NGC 7457   &$75\pm	4 $&  $62\pm  	11$  &  $15$    &  $0.9^{+0.5}_{-0.5}$\\
\hline
\end{tabular}
\end{center}

\textbf{Notes.} Column 1: galaxy name, Column 2: $\sigma_{\star}$ from \cite{Cappellari13}, measured from a single spectrum integrated within $r_{e}$, Column 3: $\sigma_{\star}$ within $r_{e}$ determined from kinematic maps of \cite{Cappellari11} with associated $1\sigma$ uncertainty, Column 4: standard deviation for the set of $\sigma_{\star}$ values averaged to determine the value in column 3, Column 5: $M_{BH}$ from the compilation of KH13. 
%\vspace{2ex} 
\end{minipage}
\end{table*}
\subsection{The AGN Sample}
To determine the best-fitting relation for AGN, VP is used in place of $M_{BH}$:
\begin{equation}\label{eq:Msigma_AGN}
\mathrm{log\:\left(\dfrac{VP}{M_{\odot}}\right)} = \alpha_{AGN}\: +\: \beta\:\mathrm{log}\left(\dfrac{\sigma_{\star}}{200\:\mathrm{km\:s^{-1}}}\right)
\end{equation}
Where, since $M_{BH}=f\:\mathrm{VP}$, $\alpha_{AGN}$ includes $\mathrm{log}\:f$.

The relation is parameterized with both error estimates (statistical uncertainty is used when a standard deviation is not available), and the best fit parameters are shown in Table \ref{tab:fits}. In general, parameterizations that include the standard deviation (column (6) of Table \ref{tab:data}) as the error in $\sigma_{\star}$ are steeper, however they are all consistent with each other. We adopt as our best-fit that which uses the statistical measurement error:
\begin{equation}\label{eq:Msigmafit}
\mathrm{log\left(\dfrac{VP}{M_{\odot}}\right)} = (7.53\pm0.26)+(3.90\pm0.93)\mathrm{log}\left(\dfrac{\sigma_{\star}}{200\:\mathrm{km\:s^{-1}}}\right)
\end{equation}
This agrees with the parameterization for AGN found by W13 ($3.46\pm0.61$), as well as for two of their quiescent galaxy subsamples: late-type galaxies ($4.23\pm1.26$), and galaxies with pseudo-bulges ($3.28\pm1.11$). It is also consistent with the quiescent galaxy parameterizations found in Section \ref{subsec:quiescent_msig}. The scatter in the relation is found to be $0.30\pm0.15\:\mathrm{dex}$, which is similarly consistent with previous studies.

Table \ref{tab:fits} also shows the best fit parameters when the alternative definition of $\sigma_{\star}$ is used. This gives a slightly shallower slope, however the fits are consistent between the two measures of $\sigma_{\star}$. 

As with previous studies, our best-fits to the AGN $M_{BH}-\sigma_{\star}$ relation are shallower than those for the quiescent sample. However this does not necessarily indicate a fundamental difference between the relations for active and quiescent galaxies. $M_{BH}$ is determined via different methods for the two samples, so different selection criteria apply. For quiescent galaxies it is necessary to resolve the gravitational sphere of influence of the BH, which is not required for AGN. Recent studies have suggested that this criterion may bias the quiescent galaxy sample, and accounting for this bias substantially reduces the discrepancy between the fits (W13, \citealt{Shankar16}).
Expanded samples of AGN and quiescent galaxies, at both the high and low $M_{BH}$ ends, are key to further investigating this difference and determining if it is physically meaningful, or simply the result of selection effects.

\begin{table*}
\begin{center}
\begin{minipage}{\textwidth}
\begin{center}
\caption{Fits to the $M_{BH}-\sigma_{\star}$ relation}\label{tab:fits}
\begin{tabular}{lccccccc}
\hline
\hline
Sample/fit& $\sigma_{\star}$& $\sigma_{\star}$ error & $\alpha$ & $\beta$  &$f$%\footnote{$log\:f$ is determined using the fit for quiescent galaxies from \cite{Woo13}, ($\beta=5.31\pm0.33$) }
& $\epsilon$\\% & $f$ \\
(1) & (2) & (3) & (4) & (5) & (6) & (7) \\
\hline
Quiescent&integrated& measurement error & $8.55\pm0.09$	& $5.32\pm0.63$ & -- & $ 0.16\pm0.06$\\
 & spatially resolved& standard deviation & $8.66\pm0.09$	& $4.76\pm0.60$ &  -- & $ 0.11\pm0.05$\\

\hline
Active & spatially resolved& measurement error & $7.53\pm0.26$	& $3.90\pm0.93$ & -- & $ 0.30\pm0.15$\\
&spatially resolved& standard deviation & $7.55\pm0.26$	& $4.00\pm0.94$ & -- &  $ 0.27\pm0.16$\\
&integrated& measurement error & $7.38\pm0.25$	& $3.53\pm0.93$ & -- &  $ 0.34\pm0.18$\\
%&literature& measurement error & $7.28\pm0.24$	& $3.48\pm0.97$ & - & - & $ 0.37\pm0.19$\\
\hline
Equation \ref{eq:Msigma_quiescent}&spatially resolved& measurement error & $7.87\pm0.15$	& $5.32$ & $4.82\pm1.67$ &$0.33 \pm0.17$\\
Equation \ref{eq:Msigma_quiescent}&spatially resolved& standard deviation & $7.86\pm0.15$	& $5.32$&$4.94\pm1.75$ &$0.27 \pm0.16$\\
Equation \ref{eq:Msigma_quiescent}&integrated& measurement error & $7.77\pm0.17$	& $5.32$ & $6.05\pm2.45$ &$ 0.43\pm0.21$\\
Equation \ref{eq:Msigma_quiescent2}&spatially resolved& measurement error & $7.73\pm0.14$	& $4.76$ & $8.49\pm2.77$ &$0.29 \pm0.16$\\
Equation \ref{eq:Msigma_quiescent2}&spatially resolved& standard deviation & $7.72\pm0.14$ & $4.76$ &$8.67\pm2.89$ &$ 0.24\pm0.14$\\
Equation \ref{eq:Msigma_quiescent2}&integrated& measurement error & $7.64\pm0.16$	& $4.76$ & $10.37\pm3.86$ &$ 0.37\pm0.18$\\
W13&spatially resolved& measurement error & $7.86\pm0.15$	& $5.31$  & $3.23\pm1.14$ &$ 0.33\pm0.16$\\
W13&spatially resolved& standard deviation & $7.86\pm0.15$	& $5.31$ &$3.27\pm1.17$ &$ 0.27\pm0.16$\\
W13&integrated& measurement error & $7.76\pm0.17$& $5.31$  & $4.03\pm1.6$ &$ 0.42\pm0.21$\\

\hline
\end{tabular}
\end{center}
\textbf{Notes. }Column 1: the first five rows give the sample being fitted, either quiescent or active, the rest show the parameterization being used, Column 2: method by which $\sigma_{\star}$ was determined, either from an average of spatially resolved spectra, or from a single integrated spectrum, Column 3: uncertainty in $\sigma_{\star}$ used in the fit, Column 4: the intercepts, Column 5: the slopes, Column 6: calculated $f$ value, Column 7: scatter in the relation.
\end{minipage}
\end{center}
\end{table*}

\subsection{The Virial Scale Factor}

We estimate $f$ by fixing the slope in equation \ref{eq:Msigma_AGN} to that for quiescent galaxies, and taking the difference between the intercepts for the two samples:
\begin{equation}
\mathrm{log}\:f=\alpha_{q}-\alpha_{AGN}
\end{equation}
Where $\alpha_{q}$ is the intercept for the quiescent galaxy sample, and $\alpha_{AGN}$ is the intercept for the AGN sample.
 
We use an adapted version of LINMIX\_ERR that allows for fixing the slope, and use the two quiescent galaxy parameterizations from Section \ref{subsec:quiescent_msig}, as well as that of W13 ($\beta=5.31$, $\alpha=8.37$), to provide a comparison with the literature. W13 is chosen because the sample of $\sigma_{\star}$ that they use contains some rotation-corrected values, so it is more consistent with our sample than others available in the literature.

The results are summarized in Table \ref{tab:fits}. The lowest value of $f$ is found with the parameterization of W13, while the highest comes from equation \ref{eq:Msigma_quiescent2}. They are all consistent with previous estimates (e.g. \citealt{Graham11,Grier13,Woo15}, W13), though our highest value is notably higher than most quoted in the literature. Our determinations are also consistent with the results of dynamical modeling of the BLR by \cite{Pancoast14}, who modeled five active galaxies (including NGC 5548 and NGC 6814) and determined $f$ separately for each, finding a mean of $\langle \mathrm{log}\:f\rangle=0.68\pm0.40$. 

For comparison we perform the same fitting using the alternative definition of $\sigma_{\star}$, which are listed in Table \ref{tab:fits}. They are generally steeper, but consistent within the errors.

As can be seen, $f$ varies significantly depending on the chosen parameterization of the quiescent $M_{BH}-\sigma_{\star}$ relation, and this is driven by the different measurements of $\sigma_{\star}$ for the quiescent sample. Given the previously-discussed issues with the $r_{e}$ determinations for the quiescent sample, it is clear that the best value to use for $f$ is still not settled. 

IFS provides more information about the galaxy kinematics, so IFS-based $\sigma_{\star}$ measurements are an improvement over previous estimates. The quiescent galaxy parameterizations presented in this work rely on such estimates, and provide a more consistent basis for comparison with our AGN sample, so $f$ values determined from these parameterizations are preferable. Equation \ref{eq:Msigma_quiescent2} should provide the best parameterization to use with the AGN sample, since the $\sigma_{\star}$ measurements do not include contributions from the rotational velocity. However the bias that arises from poorly constrained $r_{e}$ measurements for the quiescent sample does not affect the AGN sample, so they are not completely consistent. Following the discussion in Section \ref{subsec:quiescent_msig}, it is reasonable to consider that the $f$ value obtained from Equation \ref{eq:Msigma_quiescent2} is too high, and the $f$ value obtained from the parameterization of W13 is too low, so we recommend $f=4.82\pm1.67$. This is close to the median $f$ value of all those listed in Table \ref{tab:fits}, though the scatter among those values is $2.6$, which is higher than the quoted uncertainty.

\begin{figure*}
\begin{minipage}{\textwidth}
\begin{center}
\includegraphics[scale=0.95]{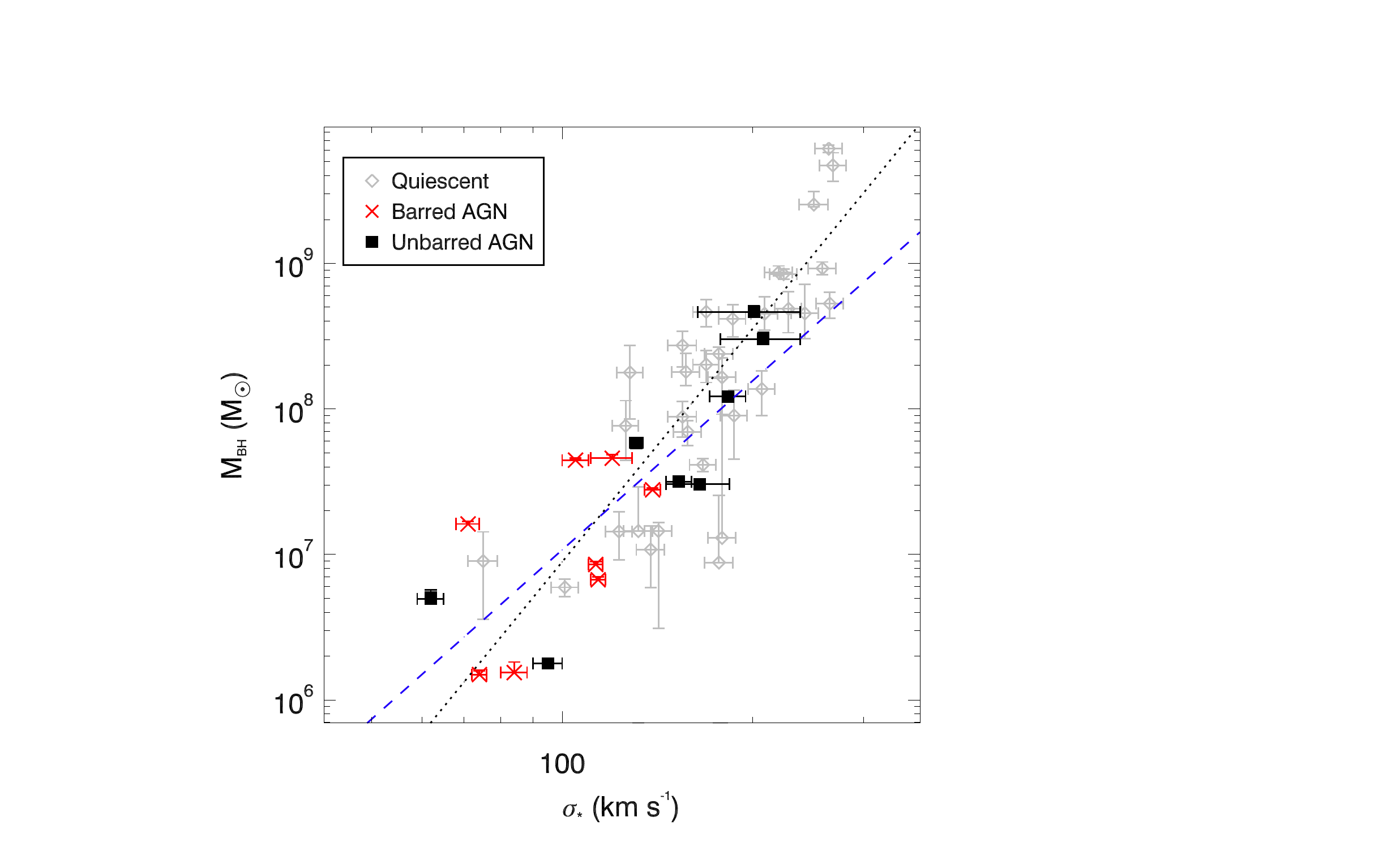}
\end{center}
\caption{The $M_{BH}-\sigma_{\star}$ relation for quiescent galaxies (grey) and AGN (red for unbarred and black for barred). The adopted best fit for the quiescent sample is shown as the dotted line, and for the AGN sample is the dashed line. VPs are converted to $M_{BH}$ using $f=4.82$.}
\end{minipage}
\label{fig:msigma}
\end{figure*}

Figure \ref{fig:msigma} shows the $M_{BH}-\sigma_{\star}$ relation, with the lines of best-fit, for both samples. While the slopes are different, the samples clearly overlap, and the absence of AGN with high $M_{BH}$ and high $\sigma_{\star}$ may well be responsible for the difference in slopes. The AGN sample is split into barred and unbarred galaxies (red and black points, respectively), since previous studies have suggested morphological dependencies in the $M_{BH}-\sigma_{\star}$ relation. We see no obvious difference between these subsamples, though we caution that this sample is too small to draw any definite conclusions. 

%\begin{figure*}
%\centering
%\begin{minipage}{\textwidth}
%\begin{subfigure}
%\centering
%\subfloat[][Plot of late type and early type]{\includegraphics[angle=90,width=0.5\textwidth]{Msigma_plot_type.eps}}
%\caption{Plot of late type and early type}\label{fig:Msigtype}
%\end{subfigure}
%\begin{subfigure}
%\centering
%\subfloat[][Plot of barred and unbarred galaxies]{\includegraphics[angle=90,width=0.5\textwidth]{Msigma_plot_bar.eps}}
%\caption{Relation for different morphological types}\label{fig:Msigbar}
%\end{subfigure}
%\end{minipage}
%\end{figure*}

\section{Summary}\label{sec:summary}
We have presented a re-calibration of the $M_{BH}-\sigma_{\star}$ relation, using $\sigma_{\star}$ determinations from IFS. Our results can be summarized as follows.

(i) Both the quiescent and AGN samples are fitted using two different definitions of $\sigma_{\star}$, and we find that including rotational broadening tends to produce a flatter slope. Our slopes are consistent with previous studies, as is the fact that fits to the AGN sample are consistently shallower than the quiescent galaxy parameterizations.

(ii) The intercepts in our quiescent fits are larger than those in the literature, due to systematically lower $\sigma_{\star}$ estimates. While this is expected for measurements from IFS, the quiescent sample suffers from poorly constrained $r_{e}$ determinations which may bias $\sigma_{\star}$ estimates low. This problem impacts the majority of studies in the literature. Larger intercepts result in larger $f$ values, demonstrating the sensitivity of $f$ to the details of the $\sigma_{\star}$ measurements. We recommend $f=4.82\pm1.67$, but caution that there remain potentially significant biases that must be addressed. 

(iii) Along with more accurate determinations of $r_{e}$, this work demonstrates the need for a significantly expanded sample of active and quiescent galaxies with $\sigma_{\star}$ from IFS, and accurately constrained $r_{e}$. This analysis clearly demonstrates that we are now in a regime where the details of the $\sigma_{\star}$ determination are important.

\section*{Acknowledgments}
MCB gratefully acknowledges support from the NSF through CAREER grant
AST-1253702 to Georgia State University. SR and MV gratefully acknowledge financial support from the Danish Council for Independent Research via grant no. DFF 4002-00275. We would like to thank the anonymous referee, who's comments have served to significantly improve the paper.

%\bibliography{ref}

\label{lastpage}
\end{document}